\documentclass[12pt]{article}
\usepackage{graphicx,color}
\usepackage{amsmath}
\usepackage{amsfonts}
\usepackage{multirow}
\usepackage{appendix}
\usepackage{url}
\usepackage{ccaption}
\usepackage{lipsum}
\usepackage{soul}
\newcommand{\beq}{\begin{equation}}
\newcommand{\eeq}{\end{equation}}
\newcommand{\beqn}{\begin{eqnarray}}
\newcommand{\eeqn}{\end{eqnarray}}

\makeatletter
\renewcommand{\p@subsection}{}
\renewcommand{\p@subsubsection}
\makeatother

\usepackage{titlesec}
\titlespacing\section{0pt}{24pt plus 4pt minus 2pt}{6pt plus 2pt minus 2pt}
\titlespacing\subsection{0pt}{12pt plus 4pt minus 2pt}{0pt plus 2pt minus 2pt}

\begin{document}
\section*{ 40 years of the Nobel prize in physics: then and now }
\centerline{R A Broglia}
%$^{1,2}$}
%\email{broglia@mi.infn.it}%\author{A. Idini $^{1}$ } %(Å£Ò»ì?) $^{1,2}$}
%\email{bortignon@mi.infn.it}
%\email{barranco@us.es}
%\email{vigezzi@mi.infn.it}
%\email{andrea.idini@gmail.com}
%\email{gregory.potel@gmail.com}

%%
%%
%%
Dipartimento di Fisica, Universit\`a di Milano,
Via Celoria 16, 
I-20133 Milano, Italy\\
The Niels Bohr Institute, University of Copenhagen, Blegdamsvej 17,
DK-2100 Copenhagen, Denmark 

%\date{\today}

\begin{abstract}
The findings for which Aage Bohr and Ben R. Mottelson became co-winners of the 1975 Nobel prize in physics provided 
the basis for a comprehensive and operative answer to the central problem in the study of the nuclear structure, namely
the identification   of the appropriate concepts and degrees of freedom that are suitable for describing the phenomena encountered.
To do so they produced a breathtaking unification of a number of well established concepts, namely liquid drop and shell models,
elementary modes of excitation, superconductivity  and quantum electrodynamics, resulting eventually in the paradigm of broken symmetry
restoration to determine  the  nuclear collective variables (CV, elementary modes of excitation): violation of translation invariance by the mean field  and 
by scattering states 
(single-particle motion), of rotational invariance  in the variety of spaces, in particular in 3D- and in gauge-space, leading to surface vibrations
and to quadrupole rotations, as well as to pairing vibrations and rotations,  with associated emergent properties 
of generalized  rigidity in these spaces, resulting from the coupling to single-particle degrees of freedom.\footnote{Invited lecture delivered on December 14th, 2015 in occasion of the celebration by
the Niels Bohr Archive of the 40th anniversary of Aage Bohr and Ben Mottelson co-sharing of the 1975
 Nobel Prize in Physics, held in the historic Auditorium A of the Niels Bohr Institute, Blegdamsvej 17, Copenhagen, Denmark} 
\end{abstract}

%%%%%%%%%%%%%%%%%%%%%%%%%%%%%%%%%%%%%%%

\section{Foreword}

The story starts in October  1974 where we, at the Institute, were waiting for the announcement of the Kungliga Vetenskapsakademien concerning the
Nobel Prize in physics. One was quite confident that this one was  the year of Aage and Ben.  However, the laureates of 1974 were 
Martin Ryle and Anthony Hewish, for their contributions to radio astrophysics\footnote{I recall that morning having met Aage at the library of the Institute, and having heard him commenting on some of the physical properties of pulsars.} (aperture synthesis technique and discovery of pulsars respectively). We were taken aback. 
Some however, decided on short call to react strongly, asking the, at the time, resident of Carlsberg's \ae resbolig, Professor Bengt Str\"omgren to make the presentation of the candidates for 1975.

\section {Then}

We all know the way things went, and on December 14th, 2015 one celebrated  within the scenario of the famous Auditorium A  of the  Institute 
of Theoretical Physics of the University of Copenhagen (now Niels Bohr Institute)  the 40th anniversary of the Nobel Prize in nuclear physics;  {\it "For the discovery 
of the connection between  collective motion and particle motion in atomic nuclei and the development of the theory of structure of the atomic nucleus
based on this connection". }

Within this context, let me quote the assessment made some years before (1972) by Phil Anderson  concerning the work which is at the basis of the event of the present
celebration: 
"It is fascinating ... that nuclear physicists (referring to A. Bohr and B.R. Mottelson, {\it Collective and individual-particle aspects of nuclear structure},
 Mat. Fys. Medd. Dan. Vid. Selsk. {\bf 27}, no. 16 (1953)) stopped thinking of the nucleus as a featureless, symmetrical little ball and realized that ... 
 it can become  (american ,RAB) football-shaped or plate-shaped. {\it This has observable consequences in the reactions and excitation spectra}  that 
 are studied in nuclear physics ... this ... research ... is as fundamental in nature as many things one might so label" (P.W. Anderson, {\it More is different.
 Broken symmetry and the nature of the hierarchical structure of science}, Science {\bf 177}, 393 (1972); within this context see App. A). And again, almost 30 years later, describing 
 the activity of the theoretical physicists in the sixties and early seventies:  "...  {\it the story of broken symmetry}  ... is a heartening story of one of those rare periods  when the  fragmentation of theoretical physics  into condensed-matter, nuclear and particle branches was temporarily  healed and we were all consciously working together 
 in exploring the many quantum consequences of the idea of broken symmetry" (P.W. Anderson, {\it A helping hand  on elementary matters}, Nature, {\bf 405}, 736 (2000);
 see App. B ).
 
 I believe that the above texts accurately  describe the {\it magic period} during which the basic concepts  and some of the fundamental results which are at the basis 
 of the 1975 Nobel prize in physics were developed at the Institute.

 {\it 2.1 The press}

 The announcement of the Nobel  prize by the Kungliga  Vetenskabakademien took place on Friday, October 17, 1975.
% That morning of Friday October 17 Aage Bohr arrived  at his office (which was in the same Kd gangen  (corridor) as mine) shortly before noon. I remember 
 %it quite distinctly, also because of two reference events. We had an appointment   to discuss on pairing, and shortly before I had received a phone call from Sven 
% G\"osta Nilsson, from Stockholm. 
 % When I saw him  I followed him  to his  office to congratulate him on the prize (I could not do the same with Ben, as he was far away, on a trip to China). After  a few minutes 
 %we heard somebody knocking at the door. The press was already there. Aage said something like: "but we had  an appointment" , to which I answered 
% that he  should better care of the journalists. 
 That Saturday, all newspapers carried the story. One of them (Aalborg Stiftende) with  excerpts of an interview Aage had  given to the press in his home in Hellerup\footnote{One is reminded of the fact that when the announcement came from Stockholm, Ben was in Bangkok in his way to China with a delegation from the Royal Danish Academy of Sciences and Letters. After his return to Copenhagen in November, Aage and Ben started coordinating and writing their Nobel lectures. I was call one day in early afternoon to Granh\o{}jen 10, to act as  sounding board and discuss specific points (Nuclear Field Theory with Ben, rotations, in particular pairing rotations with Aage). Aage sat at his desk, Ben, on the room contiguous to Aage's studio, at a low coffee table on a big, four people sofa, with his ever smiling expression. I do not know exactly why but, these two geniuses of physics appeared to me at the moment, as the kind Professor and the bright student.}. To the 
 journalist (Claus J. Deden) question
 \footnote{"whether his work is not a continuation of his father's work. The work, Niels Bohr got a Nobel prize for  in 1922."} 
 "om hans arbejde ikke er en videref\o relse af faderens arbejde.  Det arbejde, Niels Bohr  fik er Nobel-pris for i 1922" he answered (my translation): 
 "My father's work cannot be  compared with mine. His work was  an epoch-making event within the whole  development of physics.  My work is a different and 
 more modest contribution to this development. But 
  one can correctly  say, that my work  carries on that of my father ...
  %in the sense that he worked on the structure of atoms  and  came with  the first 
  %contributions for  the understanding   of the atomic nucleus. At this point,  
  we have  directly built further based on his work".
 
 The journalist then  writes that Aage emphasized, during the whole interview, how he viewed  the bestowal: "One is talking about  that 
 all, the whole group of theoretical physicists at the Institute, have taken part  in the work that is being recognized. Furthermore, 
 one refers to an international collaboration.  I cannot take all the honour for it"\footnote{At this point I cannot avoid telling you how  sad I am that Aage is not anymore among us. The best way I can remember him today, is by trying to imagine what his reaction would have been to my presentation. I could vividly see him smiling at each important statement, and mumbling something to himself. If you payed attention you could likely hear him saying: ``yes but not quite\dots''. The depth of his thoughts, the accuracy, physical accuracy, of his intuitions, made of each single issue an important piece of the whole mosaic. A never ceasing probing of the correctness of the ideas and predictions, which we sorely miss. The best antidote to it? To study his monumental opus written together with Ben when confronted with new challenges. There are still many nuggets to be extracted from this gold mine.}. 
 
% While the italics are mine, I believe that he (Aage), talking also in Ben's name, meant the above words very seriously,  as any Institute  member or visitor could experience.
 
 \vspace{5mm} 
 
% {\it 2.2 Volume II}
 % I overlapped with Aage and the family at Los Alamos in summer 1975, and remember him at the library of the Laboratory deeply concentrated in checking the
% references of Vol. II. 
% The book appeared just at the right time to carry on the dust  cover the news that the authors were co-winners of  the 1975 Nobel Prize in physics. 
% Within this context,
 %let me recall  for you a  little, charming  event, connected with this dust cover. As one can read from the last page of the Preface, Fru Hellman
% was the real driving  force behind the secretarial work. 
 %One day in early 1975  I was in Aage's office  discussing physics together with a number of visitors, among them I. Rabi (NMR). At a certain point 
 %the door opened  up and without further ado, in came Fru Hellmann who went directly  to Aage.
 %Showing him the dust cover she had just received from Benjamin, which
 %instead of red (as they had agreed) had been produced in bordeaux, she stated that such a thing 
 % was totally  unacceptable. She was quite somebody. But let us go back    to hard facts. 
 
 \vspace{5mm} 

{\it 2.2 Place and impact of the 1975 view of nuclear physics  within the general developments}
  
Let me remind you that this year (2015) we are also celebrating  another 40 year recurrence. The publication of Vol. II of Nuclear Structure, the definitive treatise on the subject 
 by Aage and Ben. 
  The 1975 Nobel Prize in physics was, to a large extent, honoring the unified vision of nuclear physics  contained in these two volumes.  Anybody who 
has studied and used them, in particular Volume II with its 748 pages, realizes that this monumental work not only testifies  to the deep vision 
Aage and Ben developed of the atomic nucleus, but the inspiration they provided  through the years  to so many practitioners, theorists  and experimentalists
alike, to explore new areas and carry out independent cutting edge research. Within this context Aage's words to the journalist concerning "a common enterprise"
is not only an expression of dignity of a unique team of scientists,  but a true description  of an heroic scientific achievement 
%\footnote{Within this context I can hardly understand that all of the work done at the Institute concerning rotational damping, is not considered at the same  level of significance, as far as fundamental research goes, to NMR as applied 
%to hard and soft matter, recognized also  by prestigious prizes. Having had to do with such a field within the context of proteins, and  having had the fortune to interact 
%with the best scientists in the field,  like e.g. the late Fleming Poulsen, my perplexity turns almost into annoyance.}
of which they were not only the central 
figures, but also the inspiring force and  the living  intuition  of this new research  field. 

\begin{figure}
%\centerline{\includegraphics*[width=.4\textwidth,angle=0]{figs/fig_8}}
\centerline{\includegraphics*[width=\textwidth,angle=0]{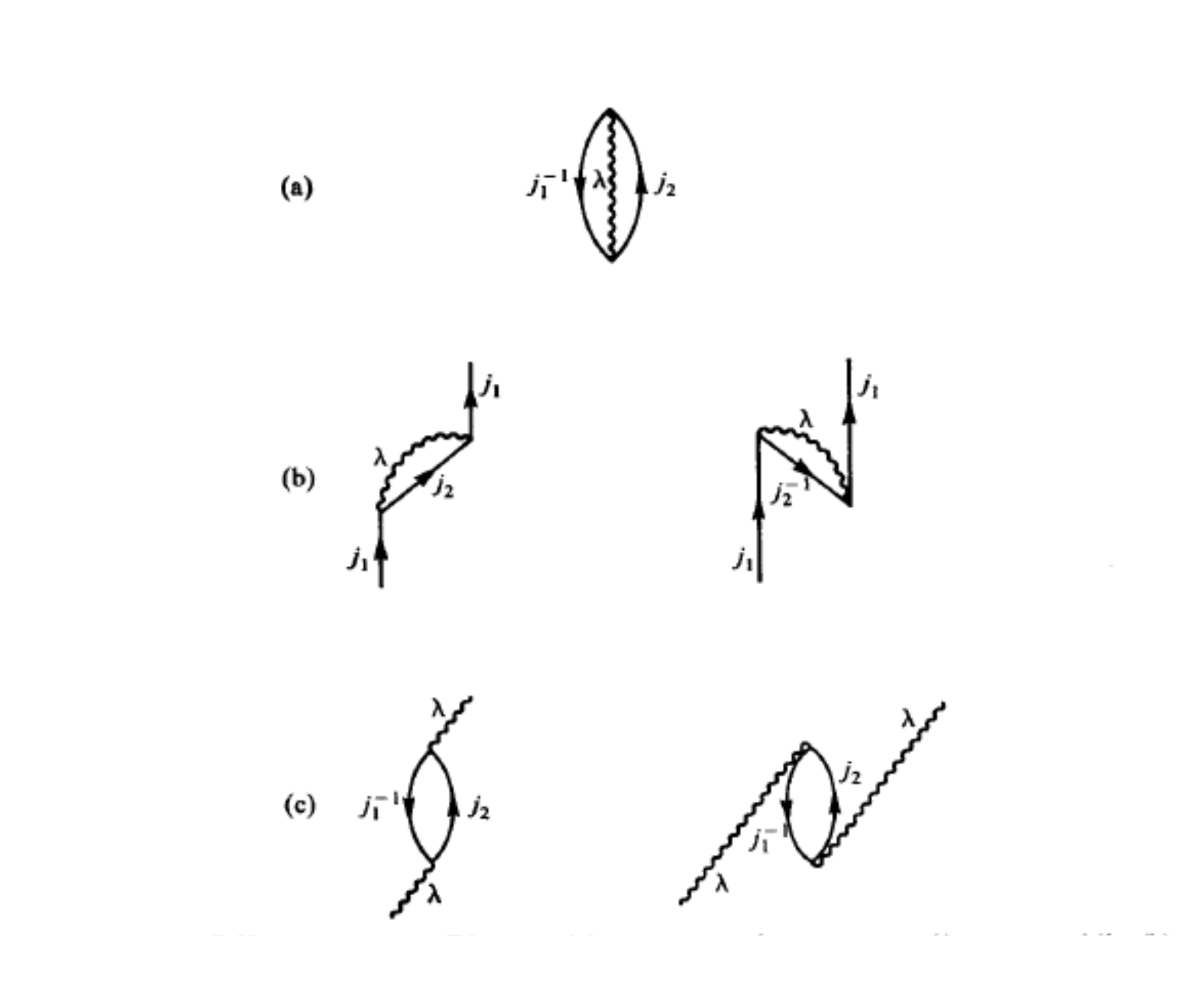}}
\caption{Reproduction of Fig. 6-11 of A. Bohr and B.R. Mottelson, {\it Nuclear Structure}, Vol. II, Benjamin, Reading (1975). Arrowed lines ($j_1,j_2$) represent particles (upward) and holes (downward). Wavy lines ($\lambda$) vibrations. In (c), the first $\lambda$ is to be understood as an external field.}\label{fig1}
\end{figure}

Returning to the first part of Aage's press interview, while 
nobody will be able to disagree with his judgment concerning the significance of Niels Bohr's work compared to that of him 
done in collaboration with
 Ben, I take issue 
at the fact that they have just built on the top of it. {\it They have changed  the nuclear paradigm in an essential way},
% As schematically 
%shown in the figure, Aage's and Ben's work provides a bridge, but also
providing  an unification  of  short, mean free path, local equilibrium models (Niels Bohr and Fritz Kalckar,
liquid drop, Niels Bohr, compound nucleus),  with a  long mean free path, independent particle motion picture, to be found
at the basis of the shell model of
Marie Goeppert Mayer and Hans Jensen\footnote{Likely, the most illustrative way to express this unification 
sounds  (quoting Ben Mottelson): ``It is a rather unfortunate  perversity of the popular terminology that regards this collective field as in some sense an antithesis to the nuclear collective effects´´. B.R. Mottelson, in Proc. of the ``E. Fermi'' Int. School of Physics, Course XV, {\it Nuclear spectroscopy}, G. Racah ed,, 
Academic Press, New York (1962), p.44. To illustrate this point to my fourth year students at the University of Milan, I used to propose the following analogy:  ``independent--particle motion in nuclei is like to drive my car from Via Celoria 16 (adress of the Department of Physics) to the Duomo and back without finding a single red light, nor another vehicle in my way''. Unarguably, the outcome of a rather coordinated city street ballet.}. In a nutshell, the particle-vibration coupling (PVC) is at the basis of the  
processes required by 
quantum mechanics concerning the evolution of systems made out of fermions and bosons (Fig.1). Namely: (a) vacuum
self-energy (Heisenberg's indeterminacy-- and Born's commutation-- relations), (b) single-particle self-energy (right, Pauli principle, Lamb shift--like diagram\footnote{A consequence of quantum mechanics at large, and of quantum field theories in particular is that the vacuum is not a simple entity. It can be virtually excited (see 1(a) and interpret $\lambda$ as the photon and the upward (downward) going particle as the electron (positron)). To the question of Rabi of whether the polarization of the vacuum could be measured (A. Pais, The genius of Science, Oxford (2000)), Lamb gave a quantitative answer, both experimentally and theoretically (W. E. Lamb Jr. and R. C. Retherford, Phys. Rev. \textbf{72}, 241 (1947), N. M. Kroll and W. E. Lamb Jr. \textbf{75}, 388 (1949)). In the nuclear case Fig. \ref{fig1}(a) represents the ground state fluctuations while graphs (b) result by adding one nucleon to the system. The  second order process associated with antisymmetrization between the single nucleon considered explicitly and the nucleons out of which the vibrations are built, as well as the symmetrization between nuclear vibrations renormalizing the nucleons and eventually excited by external fields were first introduced in B. R. Mottelson (Proc. of the Int. Conf. on Nuclear Structure, Suppl. to J. Phys. Soc. Japan, \textbf{24}, 87 (1968); see also A. Bohr and B. R. Mottelson Nuclear Structure Vol II (1975), p. 427, Fig. 6.10; see also D. R. B\`es, Special Physic Scripta Edition --40 year anniversary-- Nobel Prize in Physics 1975). Within this context one is reminded of the fact that Aage attended the Pocono conference (30/3--1/4, 1948) where Feynman presented his version of QED (after Schwinger). In explaining his formalism, he pointed out that one did not need  to worry about the Pauli principle in intermediate states, as there were diagrams which properly took  care of it. Likely, the fact that to Teller's question:``You mean that helium can have three electrons in the $s$--state for a little while?'' Feynman gave an affirmative answer with the ensuing chaos (see Schweber QED Princeton Univ. Press 1994, p. 442) arguably could also have been at the basis of Ben Mottelson contribution to the Tokyo conference, also with the input of Aage. ``Major parts of the present report have been taken from a monograph which is in preparation jointly with Professor A. Bohr'' reads in the acknowledgment of Ben's Tokyo contribution (appeared also as No. 265 of NORDITA publications).};
left, time ordering). These processes allow, for nucleons moving several MeV away from the Fermi energy,  to undergo 
  transitions (damping)  between independent--particle motion with a  long mean free path towards a regime of short mean free path 
(as compared to nuclear dimensions), through the variety  of doorway states.  That is, states containing
an odd number of fermions and any number of collective vibrations (bosons). 
Scattering (PVC) vertices and their consequences cannot be avoided, not even within RPA
(Bohm and Pines, see diagram (c) Fig. 1), in keeping with  Pauli principle and 
with the fact that both even or odd number of fermions can be present in the  systems under study.

The work  of Aage and Ben was also instrumental to   establish  
connections between nuclear structure with  many--body and field theory physics:  Elementary  modes of excitation
(Lev Landau); Spontaneous symmetry breaking   (BCS, Josephson, Phil Anderson);  Field theories (Richard Feynman).

The validity of the ensuing  unification in terms of the symmetry breaking  restoration 
paradigm\footnote{It is of notice that the paradigm of broken symmetry  restoration
which most physically and economically allows  one to introduce and determine the elementary modes of excitation , is not to be found in the 1953 paper, nor for that sake in 
A. Bohr's contribution to the 1964 Paris Congress (A. Bohr, Comptes Rendus du Congr\`es International de Physique Nucleaire, Vol. I,  I.P. Gugenberg ed. ,
Editions du Centre National de la Recherche Scientifique, Paris, (1964), p.487), while it is embodied in the discussion of pairing rotations and vibrations of 1968 supplied by
the corresponding references  (A. Bohr, 
{\it Pair correlations and double transfer reactions}, Nuclear structure,  Dubna Symposium, IAEA (1968) p.179).} 
and resulting elementary modes of excitation and couplings , which looks short of unbelievable at first sight, 
is likely to be deeply rooted in the fact that many, if not most, high-dimensional models, as well as real processes, are "sloppy", as a consequence 
of the emergence of variables of greatest  interest \footnote{Cf. M. Buchanan, {\it Wheat from the chaff}, Nature Physics, {\bf 11}, 286 (2015);  M.K. Transtrum et al.,
J. Chem. Phys. {\bf 143} 010901 (2015).}
% (Collective variables, CV).

\vspace{5mm}

% {\it 2.4 The Varenna school of 1976}

%The 69th "International Enrico Fermi School of Physics" of 1976 which had been arranged much before the announcement of the Swedish Academy, was, as can 
%be schematically seen from the figure,  a remarkable 
%event, where  much of the physics at the forefront of nuclear  research was discussed. 

%\vspace{5mm}

%{\it 2.5 Open problems, and shortcomings}
   
% From the above one can be led to the impression that the story, as recounted in Vols. I and II was all right , and only details were missing. 
 %Not quite. Let me give an example. 
 % The optical potential, Sect. 2.4-c  of Vol. I p. 213. In the discussion at p.216 concerning the role of the surface in the absorptive 
 %process only inelastic channels are referred to. We know that   particle transfer channels can be, as a rule, the dominant ones. 

\section{Now}

The relevance for today's nuclear physics  research of Aage's and Ben's contributions recognized by the Swedish Academy
can be exemplified by quoting from their Nobel lectures. 
In p. 370 of Rev. Mod. Phys. {\bf 48} (1976), Aage writes: "The condensates in superfluid
systems involve a deformation of the field that creates the condensed bosons or  fermion pairs. Thus the process of addition 
or removal  of a correlated  pair of electrons from a superconductor (as in a Josephson junction) or 
of a nucleon pair  from a superfluid nucleus constitutes a  rotational mode in which particle  number plays the 
role of angular momentum (Anderson, 1966). Such pair rotational spectra, involving a family of states in different nuclei,
appear as a prominent feature in the study of two-particle transfer processes 
(Middleton and Pullen , 1964; see also Broglia et al, 1973)\footnote{P. W. Anderson, Rev. Mod. Phys. {\bf 38}, 298 (1966); 
R. Middleton et al., Nucl.Phys. {\bf 51}, 77 (1964);
R. A. Broglia et al., Adv. Nucl. Phys. {\bf 6}, 287 (1973).}.
The gauge space 
is often felt as an abstract construction but, in the particle-transfer process\footnote{The fingerprint of spontaneous symmetry breaking in finite, many--body systems, is the presence of rotational bands associated with symmetry restoration. To qualify as a rotational band, a set of levels must display enhanced transition probabilities (absolute cross sections), associated with the operator having a non--vanishing value in the (degenerate) ground state (order parameter). In the present case (pairing rotational bands), of the two--nucleon  transfer operator. In other words, cross talk (absolute transfer cross sections) between a member of a pairing rotational band and states not belonging to it, should be much weaker than that between members of the band. It could be argued that also important for the characterization of a pairing rotational band is the parabolic dependence of the energy with particle number. True, but many non--specific effects can modify this dependence, without altering the gauge kinship (common intrinsic state).} it is experienced  in a very real manner".
While the sequence of levels displayed in Fig. 2  have been known for quite some time, 
the account of the absolute two-nucleon transfer differential cross section 
%displayed in Fig. 5  of 
is of new date\footnote{Potel et al. Rep. Prog. Phys. {\bf 76} 106301 (2013)} , making 
Cooper pair transfer a quantitative probe of pairing correlations in nuclei, carried out in terms of absolute cross sections and not relative ones as done before. A result which  took short of forty years to be accomplished 
and involved an important fraction  of the nuclear physics community (see Fig. 10 of Potel et al. (2013)). 
Paramount in this development was the role played by Claus Riedel at the beginning and that
of Aage Winther and Ben Bayman throughout. 

Let me now quote from Ben's Nobel lecture (Rev. Mod. Phys. {\bf 48} (1976), see p. 382):
"As illustrated by these examples, it appears that the nuclear field theory based upon
the particle-vibration coupling provides a systematic method for treating the old problems
of the overcompleteness of the degrees of freedom,
as well as those arising from the identity of the particles
appearing explicitly and the particles
participating in the collective motion (Bes et al., 1974; Bohr and Mottelson, 1975)\footnote{D. R. Bes et al., Phys. Lett. B {\bf 52}, 253 (1974);
A. Bohr and B. R. Mottelson, {\it Nuclear structure}, Vol. II, Benjamin, Reading (1975)}.
This development is one of the active frontiers in the current exploration
of nuclear dynamics" (see  App. C). Fig. 3, which displays  an example  of NFT at work  today\footnote{A. Idini et al., Phys. Rev. C {\bf 92}, 031304(R)  (2015)}, provides a complete  characterization  
of $^{118,119,120,121,122}$Sn, and testifies to the actuality 
of Ben's words. 
This result has again taken  many decades to be obtained and has involved an even wider number of contributing groups  and practitioners.
Among them  one has to mention that of Pier Francesco Bortignon  and the towering one of  Daniel R. Bes.

  Within this context, a textbook example of the fact that shell and liquid drop models are to be treated on equal footing, and eventually melt together, 
in the description of the nuclear structure,
is provided by parity inversion in $_3^{11}$Li. Due to processes like those shown in Fig. 1(b), 
the level sequence $0p_{1/2},1s_{1/2}$ is inverted, the magic number N=8 melting away to lead to N=6 as new magic number
\footnote{F. Barranco et al, Eur. Phys. J. A {\bf 11}, 385 (2001).}. In other words, self-energy effects are, as large, or even larger, than mean field effects in this exotic halo nucleus. Furthermore, the binding of the halo Cooper  pair to the closed shell system (core $_3^9$Li$_6$) is essentially due to the induced pairing $v^{ind}_p$  arising 
from the exchange of collective vibration between the two neutrons, 
the contribution of the strong nucleon-nucleon bare interaction $v^{bare}_p$ 
to the Cooper pair binding being, in the present case, quite small. Such a possibility ($ v^{ind}_p > v^{bare}_p$), had been already
envisaged \footnote{In spite of this, Ben was rather negative
concerning the whole issue of the induced pairing interaction in nuclei, arguing that most of the associated
effects, were to be included in the polarization of the mean field, an understanding which is to be found at the basis of the pairing plus quadrupole (1959) Bohr and Mottelson model, according to which the long range part of the nuclear interaction is responsible for the creation of the average field, the short range part leading mainly to pairing correlations. Be as it may,  the paper Barranco \textit{et al} (2001) (footnote 15), originally submitted to 
Physical Review Letters, was rejected by the referee who not customary signed his report: Ben R. Mottelson. 
 The predictions   of the paper in question were eventually confirmed and provided, for the first time, evidence for phonon mediated 
pairing in nuclei 
(Tanihata et al, Phys. Rev. Lett. {\bf 100}, 192502 (2008); cf also Potel et al, Phys. Rev. Lett. {\bf 105}, 172502 (2010)).
Arguably, the above story just tells one simple thing. If you are convinced of your ideas and results, just publish them. There is a delicious anecdote related to the referee report. In our answer to it, we appended a four pages long discussion where many of the consequences of the interweaving of single--particle and collective vibrations where derived in detail, within the framework of a simplified model. Our referee was very happy with it, and suggested to use it as appendix to our Phys. Rev. Lett., letter which already was slightly longer than the four allowed pages (the appendix was eventually published in F. Barranco et al. Phys. Rev. C \textbf{72}, 0543149 (2005).} in Vol. II (cf. last paragraph  in Sect. 6-5f, p. 432).

 If there was need for any further  documentation of the soundness of the Nobel Committee 1975 decision, 
 one can refer to the consequences  the paper by A. Bohr,
 B.R. Mottelson and D. Pines (A. Bohr \textit{et al}, Phys. Rev. \textbf{110}, 936 (1958)), introducing BCS pairing in nuclear physics\footnote{Pairing in nuclei have been introduced in nuclear physics three times. The first two in terms of energy arguments and thus not exactly specific ones (enhanced stability of even--even systems compared to odd--even, even--odd, odd--odd (see e.g. W: Heisenberg Die Physik der atomkerne Friedr. Vieweg and Sohn de Bramschweig (Berlin) 1943); presence of a gap in the low--energy spectrum of deformed nuclei, Bohr, Mottelson and Pines (1958)). The third time in terms of the specific probe, namely two--nucleon transfer reactions (S. Yoshida, Nucl. Phys. \textbf{33}, 685 (1962), A. Bohr, Paris congress, see footnote 8).} has had over the years,
  by referring to the contributions collected in the 670 pages volume
 {\it  Fifty years of nuclear BCS.}  
 \footnote{{\it  Fifty years of nuclear BCS},
 R.A. Broglia and V. Zelevinsky eds., World Scientific, Singapore (2013).}
 
 \section{Conclusions}
 
 The unification of single-particle and collective motion  which is at the basis  of the ground-breaking contribution   Aage Bohr and Ben R. Mottelson 
made to nuclear physics 
inspired  and helped in a major way at developing the paradigm of broken symmetry restoration, and   the physical tools to implement structure calculations in terms of the 
particle-vibration coupling mechanism (Feynman diagrams).
Going beyond the harmonic approximation through 
scattering vertices, it allows for a systematic description of anharmonic effects and (doorway) damping. These  major
scientific breakthroughs were  
properly recognized by the Nobel committee in 1975.

Volume II of their treatise on Nuclear
Structure   is certainly no easy reading, 
in keeping with the fact that aside from containing  the material describing the above mentioned phenomena 
and their implementation in concrete cases, it also provides a comprehensive account 
of the contributions of extensive NBI and international collaborations, and the basis to compare the theoretical physical results with the 
experimental findings obtained  making use of nuclear reactions and decay processes. The fact that a three volume project became, in the end, a two volume one 
did not help on easiness of reading. Nonetheless, practitioners of other fields of physics, like condensed matter, cannot complain, neither explain that 
their knowledge of nuclear physics is  limited at best, because nuclear physicists express themselves in a complicated language\footnote{Within this context one can agree that the language of nuclear filed theory and of its applications as developed by Aage, Ben and collaborators, was particularly original and for sure non--standard (see, e.g., the review of G. Breit and G. E. Brown of Volume II of Nuclear Structure, Pysics Today, March 1977 pp. 59--62). This is in fact, one of the main challenges awaiting to be met. Namely, that of establishing a direct connection of the physics which is at the basis of Nuclear Structure Volume I and II, and thus of the 1975 Nobel prize in physics, with the theory of many--body physics and of effective field theories, heavily used by the physics community. The eventual outcome could prove momentous not only for present--day nuclear research, but also for finite many--body physics at large.}. Unless physical insight  and sheer  unrelenting 
use of physical intuition \footnote{Quoting: "A priori one would believe that the easiest thing is to communicate 
simple concepts and phenomenology. This is totally wrong. Students are accustomed to mathematics, and the thing that they understand
are mathematical models (JRS)", and : "On the other hand, people have to grow up (PWA)". From a discussion
with Bob Schrieffer (JRS) and Phil Anderson (PWA) which took place in the first half of July 1987, at the shadow 
of the giant  magnolia of Villa Monastero, Varenna , Como Lake (see R.A. Broglia and J.R. Schrieffer, eds., International School of Physics ``Enrico Fermi'', Frontiers and Borderlines
in many-particle physics,  Course CIV, North Holland, Amsterdam (1988)).} can be  deemed too trying.
% write too complicated [questa frase non la capisco bene...]. 
Here (volumes  I and II) you find fermion 
superconductivity and Josephson pair tunneling and its particularization to finite systems  and its extension to neutron (star) matter (pulsars).
Furthermore spontaneous breaking of symmetries  and associated Goldstone modes\footnote{Quoting from Aage's Nobel lecture ``\dots the very occurrence of collective rotational degrees of freedom may be said to originate in a breaking of rotational invariance, which introduces a ``deformation'' that makes it possible to specify an orientation of the system. Rotation represents the collective mode associated with such a spontaneous symmetry breaking (Goldstone boson)''. Discussing with Aage when applying these concepts to the treatment of pairing rotational bands (R. A. Broglia \textit{et al} Phys. Rep. \textbf{335}, 1 (2000)), I remarked that the energy of the corresponding Goldstone mode should approach zero linearly as $N\rightarrow 0$, while the energy of the members of a pairing rotational band are quadratic in $N$.``True, but in fact one has to calculate this energy in the laboratory system, where one can measure it. For this purpose one should add the Coriolis--like term, linear in $N$''. That simple.}, let alone all the 
elements, coupling constants and numerical factors to calculate, predict and analyze observables, within the remarkable femtoworld of the atomic nucleus.  

 Discussions with Gregory Potel, Francisco Barranco, Enrico Vigezzi, and Andrea Idini concerning a number
 of "now" subjects are acknowledged. A similar acknowledgment is extended to Pier Francesco Bortignon, also concerning 
 a number of "then" subjects.

\newpage
\setcounter{equation}{0}
\appendix   
\chapter{\it Appendix A. Good physics and reductionism}
 
 Let us quote from Leon Cooper's contribution to the volume \textbf{BCS: 50 Years} \footnote{L. N. Cooper, BCS: 50 Years, eds. L.N. Cooper and D. Feldman, World Scientific, Singapore (2012), p.12}:  "It has become fashionable ... to assert .. that once gauge symmetry
 is broken the properties of superconductors follow ... with no need to inquire into the mechanism by which the symmetry is broken. This is not ... true, since broken gauge 
 symmetry
 might lead to molecule-like pairs and a Bose-Einstein rather than BCS condensation ... in 1957, we were aware that what is now called broken gauge symmetry would, under some circumstances (an energy gap or an order parameter), lead to many of the qualitative features of superconductivity .. the major problem was to show how an energy gap,
 an order parameter  or ``condensation in momentum space''  could come about ... to show 
 ...how the gauge--invariant symmetry of the Lagrangian could be spontaneously  broken due to interactions which were 
 themselves gauge invariant"..
 
Let us now go to the last contribution , namely that of Steven Weinberg (p.559). To do so, we have  to browse through p. 85 where among other things, 
small ``technical details''\footnote{A sheet of white paper on which a large rectangle had been drawn, with few written lines. Above:``This is to show the world that I can paint like Titian''. Below:``Only technical details are missing''. This was Pauli's comments to the US journalists (he was visiting the States at that time), as a reaction to the german press which had reported, likely not completely correct, on Heisenberg's G\"otingen lecture of February 24th, 1958, presenting his new field equation, partially worked out together with Pauli himself, which allegedly provided a unified description of elementary particles (A. Hermann, Heisenberg 1901--1976, 1976 Inter Nations, Bonn--Bad Godesberg).} concerning superconductivity
 in metals (namely electron-phonon coupling), are discussed by David Pines.
 
 Quoting from Weinberg: " Most of us do elementary particle physics ... because we are pursuing a reductionist vision... I think that 
 the single most important thing accomplished by ... (BCS) was to show that superconductivity is {\it not} part of the reductionist frontier ... (but) .. nothing
 more than a footling small interaction between electrons  and lattice vibrations ... All of the dramatic exact properties of superconductors ...
 follow from the assumption  that electromagnetic gauge invariance is broken ... with  no need to inquire into the mechanism by which 
 the symmetry is broken .. their (BCS, 1957) attention was focused on the details of the dynamics rather  than the symmetry breaking".
 
 This very last sentence egregiously summarizes  one of the facets of Aage's and Ben's contribution to nuclear physics, namely how to go beyond 
 symmetries and become able to calculate observables  which provide an 
 account  of the measured values  within experimental errors 
 \footnote{At a dinner in a downtown  Copenhagen steakhouse (situated in \AA benr\aa $\,$ street), Aage Bohr remarked to his two young guests,  that describing the properties  of a many-body system like the 
 atomic nucleus in terms 
 of symmetries, can provide   important insight into the structure  
 and dynamics of the system. On the  other hand,  when one is able  to describe  the same 
 system in terms of the detailed motion of the particles (nucleons), and of the 
 fields that act upon them is that one obtains, as a rule, 
 a true  understanding of the system under consideration.}
 \footnote{Within this context one can mention the unique mechanism to break gauge invariance in nuclei and bind the neutron halo Cooper pair of $^{11}_3$Li$_8$ (cf. footnotes 15 and 16).}.
  \vspace{5mm}
  \newpage
\setcounter{equation}{0}
\appendix    {\it Appendix B. The spontaneous symmetry breaking restoration paradigm in nuclear structure and reactions} 
  
\makeatletter
\renewcommand{\theequation}{B\@arabic\c@equation}
\makeatother

  The restoration of translational invariance, and of rotational invariance both in 3D-- and in gauge-space, spontaneously broken
  in nuclei by the privileged position of the finite system (e.g. of its center of mass), by surface deformations, and by superfluidity  respectively, 
  leads to zero frequency modes ($\omega \to 0$) of vanishing restoring force. Thus, divergent  zero point fluctuations 
  (ZPF), but finite inertia  namely $Am, \hbar^2/2{ \cal J}$ and $\hbar^2/2 {\cal J}_p$, where $A$ is the mass number, $m$ the nucleon mass,
  ${\cal J}$ and ${\cal J}_p$ the moments of inertia in 3D-- and in gauge--space. The resulting elementary modes of excitation  are particle motion, and rotations
   and vibrations. In particular, quadrupole rotations and surface vibrations, and pairing rotations and  vibrations.
   
 Because all of the nuclear degrees of freedom are exhausted  by the particle degree of freedom, single-particle motion displays a finite overlap 
 with the variety of rotations and  vibrations, leading to couplings bilinear in fermions and linear in bosons. In particular, 
 to the particle-vibration coupling ($\Gamma^{\dagger} (\beta=0) a^{\dagger}_{\nu} a_{\nu}$,
 $\Gamma^{\dagger} (\beta=-2) a^{\dagger}_{\nu} a^{\dagger}_{\bar \nu}$ and $\Gamma^{\dagger} (\beta=+2) a_{\bar \nu} a_{\nu}$,  where 
 the transfer quantum number is $\beta=0$ for surface modes and $\beta= \pm 2$ for pairing modes, see Fig. 4).
 
 Potential energy privileges fixed position between particles. Fluctuations, classical or quantal, symmetries. Regarding particle 
 motion, such  competition is embodied in the quantality parameter
 \begin{equation}
 q = \frac{\hbar^2}{m a^2} \frac{1}{|v_0|},
  \end{equation}
  where $m$ is the nucleon mass, $v_0$ and $a$ the strength and the range of the strong NN-potential ($v_0=-100$ MeV, $a \approx $ 1fm). The
  above equation thus provides the ratio 
  between the kinetic energy of confinement, and  the potential  energy. Because $q \approx 0.4$, nucleons in the nucleus 
  are delocalized, and mean field is  a good approximation. 
  
  In the case of independent pair motion , pairs of nucleons moving in time reversal states are correlated over distances $\xi \approx \hbar v_F/(2|E_{corr}|) 
  \approx$ 20 fm ($v_F/c \approx 0.3$; $E_{corr} \approx -1.5 $ MeV in the case of normal nuclei, and equal to the pairing gap  $\Delta \approx 1.4$ MeV  in the case  of superfluid nuclei). The value of the associated 
  generalized quantality parameter 
  \begin{equation}
 q_{\xi} = \frac{\hbar^2}{2 m \xi^2} \frac{1}{|E_{corr}|} \approx 0.03,
  \end{equation}
  testifies to the fact that  nuclear Cooper pairs partners are  solidly anchored to each other, and thus decoupled from the mean field\footnote{This fact has, among other things, important consequences for the moment of inertia $\cal J$ of e.g. quadrupole rotational bands associated with the restoration of spontaneous symmetry breaking of rotational invariance resulting from deformations in 3D--space (see P. W. Anderson, More is Different Sect. 2). In fact, the observed moments of inertia are considerably smaller than the rigid moment of inertia $\mathcal J_{rig}$, typical of independent particle motion ($\mathcal J\approx\mathcal{J}_{rig}/2$). On the other hand, the observed values of $\cal J$ are about a factor of 5 greater than the irrotational moment of inertia $\mathcal J_{irrot}$, typical of a liquid drop of the corresponding shape. This is the scenario of deformed mean field (Nilsson model) in which nucleons form extended Cooper pairs (A. Bohr and B. R. Mottelson, Nuclear Structure, vol II (1975) p. 75).}. Cooper pair transfer, 
  which mainly proceeds successively, probes pairing correlations equally well than simultaneous transfer.\footnote{To assert the contrary
  will be similar,  within the quantum language, to posit  that the two slit experiment breaks, when the slids are far away, the photon in two.} 
  
  The particle-vibration couplings displayed in Fig. 4 clothe the single-particle motion. Thus, a complete 
  characterization of physical single-particle states requires not only single-nucleon transfer but also inelastic and two-particle transfer. 
  This is also true to probe pairing correlations, in keeping with the fact that Cooper pairs are made out of physical particles, bound not only by the NN--$^1S_0$ bare
  potential ($a\approx 1$ fm), but also  by the exchange of phonons, of wavelength $\lambda$ of the order of the nuclear dimensions ($\lambda=2\pi R/L $ larger or equal than $R$, $R$ being the nuclear radius and $L$ (2--5) the phonon (vibration) multipolarity). Within this scenario, one can posit that pairing correlation in nuclei receive equally important contributions from short-- (bare--) and long-- (induced--) interactions. 
  
  \vspace{5mm} 
  \newpage
\setcounter{equation}{0}
\appendix   
  {\it Appendix C. The well funneled nuclear structure landscape}
  
  Figure 3  displays the root mean square deviations $\sigma$ between the value of the theoretical predictions and a "complete" set of experimental  observations (results) which exhaustively  characterize the open-shell
  superfluid nucleus $^{120}$Sn, and involve the island of isotopes $^{118,119,120,121,122}$Sn, namely (cf. ref. in  footnote 14)): 
  
  a)   Coulomb excitation and subsequent $\gamma-$decay ($^{119}$Sn)
  
  b) One-particle transfer reactions ($^{119,121}$Sn)

  c) Two-particle transfer reactions ($^{118,120,122}$Sn), 
  
calculated as a function  of a number of quantities associated with  single-particle and collective motion.
%\footnote{That is, making use  of SLy4 effective interaction (mean field) and $v_{14} ( ^1S_0)$ Argonne pairing NN
%potentials, one is able to reproduce  within 10\% accuracy the experimental data with just three parameters, namely $\kappa_2,\kappa_3$ and $\delta \epsilon_{d_{5/2}}/
%(\epsilon_{d_{5/2}} - \epsilon_F) \approx 0.17$. The parameters $\kappa_2$ and $\kappa_3$ are strongly constrained by the data.}
% $m_k, v_{14}, k_2$ and $\delta \epsilon_{d_{5/2}}$. 
 One expects  that the variety of $\sigma-$values do display a minimum for the set of physical quantities determining, through the  interweaving of  the elementary modes of excitation, the physical states which reproduce the data, regardless  of the properties 
 one is looking at or the probe one is using to do it. In other words a well funneled nuclear structure landscape,
 as is duly observed.

 Let us return to "More is different" of Phil Anderson (Sect. 2). I quote: ``Three or four  or ten (or for that sake  $^{120}_{50}$Sn$_{70}$, RAB) nucleons whirling about each other do not define ("a condensate" or a "vibrating surface" RAB)... It is only 
 as the  nucleus is considered  to be a many-body system - in what is often called  the $N \to \infty$ (thermodynamic, RAB) limit\footnote{Within this context ($N\rightarrow\infty$), in a number of occasions during November 2015, I discussed with Ben on the structure of the halo of $^{11}$Li. In connection with the fact that one was trying to understand a mechanism to break gauge invariance in a situation very far away from $N\rightarrow\infty$, as we were dealing with a single Cooper pair, he asked a number of times, almost to himself, what is large $N$? From his smile I could guess that he knew the answer, also very well. To my question of what he meant by that he said: ``well, is five or ten a large number, or a million, or for that sake Avogadro's number?''. The reader is reminded of the fact that 5--10 is the number of Cooper pairs of a ``normal'' superfluid nucleus, while $10^6$ is the number of Cooper pairs of a low--temperature ``normal'' metallic superconductor, whose center of mass fall within the extent of a giving pair function (correlation length). In this connection see also P. W. Anderson, More is Different, Sect. 2.} - that such behaviour is rigorously definible 
 \footnote{For example, a swing will  have  a very simple and perfectly funneled landscape.} . We say to ourselves: ``A macroscopic body of that shape  would have  such--and such-- a spectrum of rotational and vibrational excitations''(both in 3D-- and in 
 gauge--space, and a well funneled behavior with respect to  variations of the $k$-mass, the strength  of the electron--phonon coupling $\lambda$ (mass enhancement 
 factor, $m_{\omega} = m(1+\lambda)$), the Coulomb screening $\mu$, etc., RAB). 
 
 When we see such a behaviour  in the nuclear case (center diagram of Fig. 3), 
 even not so well defined, and somehow imperfect, we recognize that the physical collective variables CV (Local Elementary Modes of Excitation), 
 defined making use of concepts inspired and worked out to  large extent by Aage Bohr and Ben Mottelson, and embodied  in the spontaneous symmetry breaking  restoration 
 paradigm, provide an accurate (within 10\% error) and economic description  of the nuclear structure as probed by nuclear reactions. Also the background 
 on which to build upon, to interpret  and predict new experiments, 
 in particular concerning exotic nuclei. 
 This is, unarguably, the great achievement  that the 1975 Nobel prize in physics recognized.
  
 \newpage
  
  \begin{figure}
%\centerline{\includegraphics*[width=.4\textwidth,angle=0]{figs/fig_8}}
\centerline{\includegraphics*[width=0.82\textwidth,angle=0]{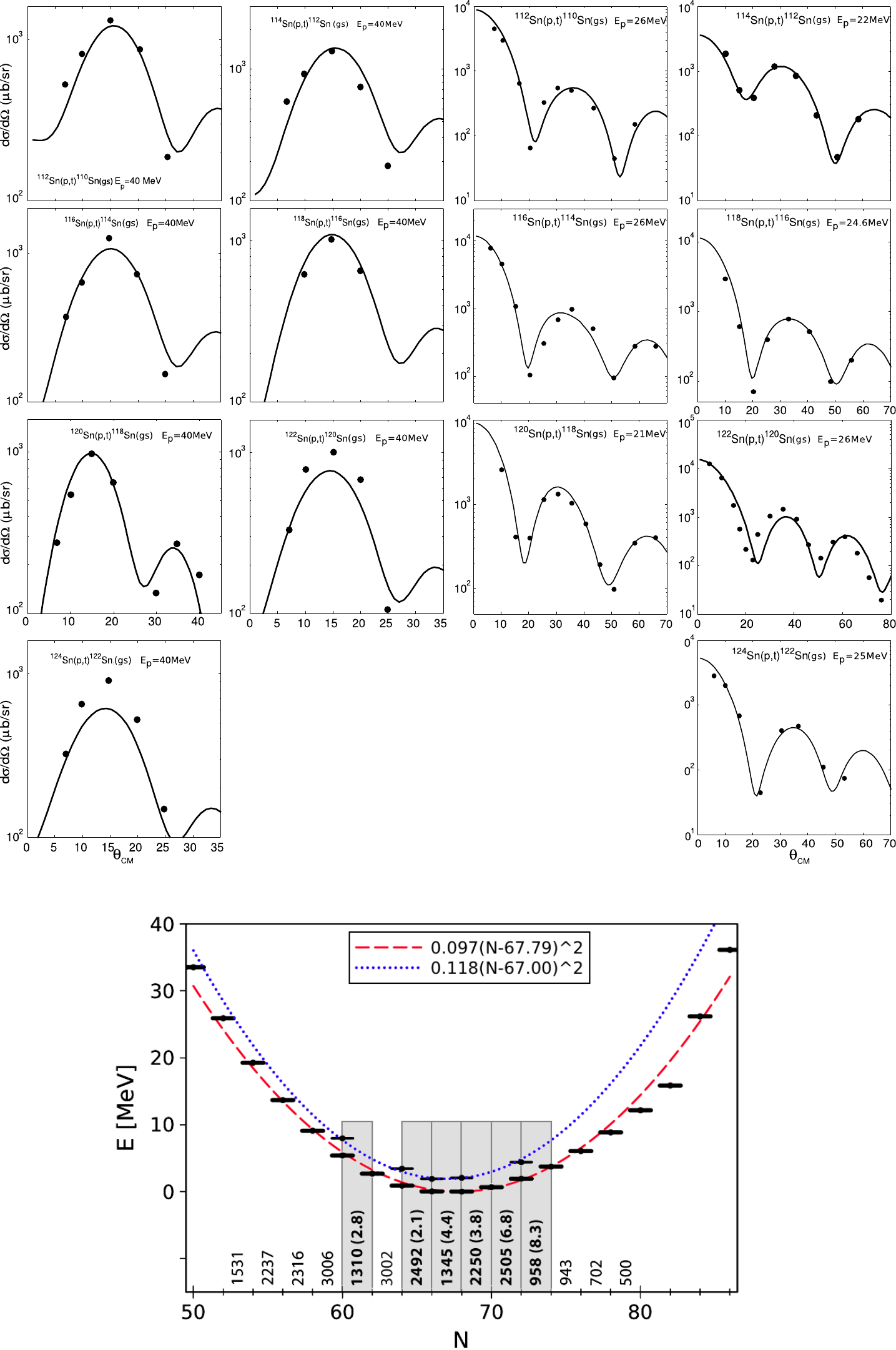}}
 \caption{Calculated (continuous curves) absolute differential cross sections associated with the reactions $^A$Sn(p,t)$^{A-2}$Sn (gs) 
 carried out at 40 (left) and 
26 (right) MeV, in comparison with the experimental findings (solid dots). At center bottom, 
the energies of the ground state, and excited states, pairing rotational bands (for details see footnote 12 and ref. therein).}\label{fig2}
\end{figure}
 
  \begin{figure}
%\centerline{\includegraphics*[width=.4\textwidth,angle=0]{figs/fig_8}}
\centerline{\includegraphics*[width=1.2\textwidth,angle=90]{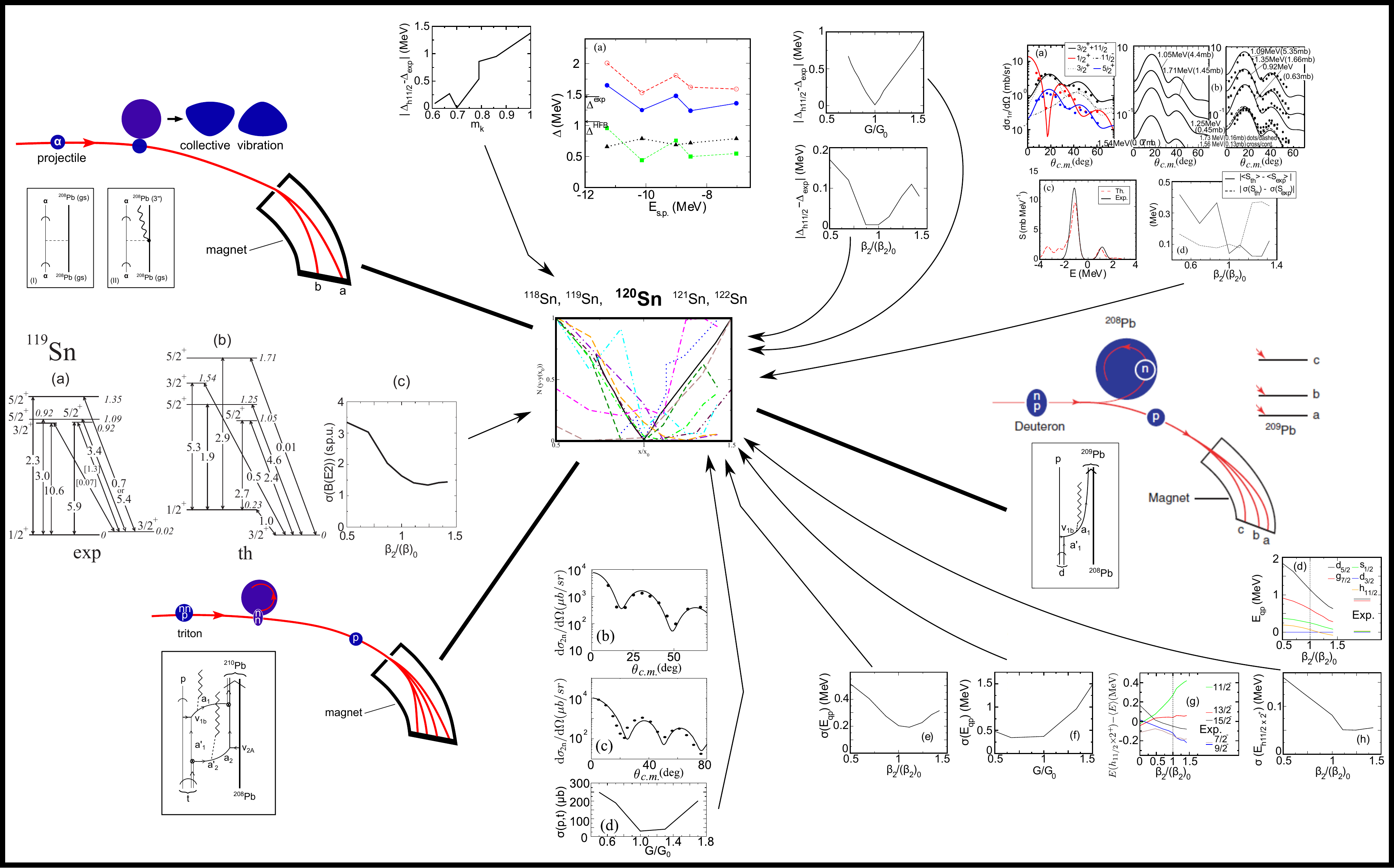}}
\caption{Results of the observations using $(\alpha,\alpha')$ followed by $\gamma-$decay, (d,p), (p,d),(t,p) and (p,t)
reactions, which completely characterize   the nucleus $^{120}$Sn and involve the island 
of superfluid nuclei $^{118,119,120,121,122}$Sn. By varying the value of a number  of inputs (effective $k-$mass,
pairing coupling constant, collectivity of the quadrupole vibration, etc .) one observes (center boxed diagram) that  the nuclear landscape is well funneled (for details see ref. in footnote 14).
%cf. A. Idini et al., 
%Phys. Rev. C {\bf 92}, 031304 (R) (2015).
}\label{fig3}
\end{figure}

  \begin{figure}
%\centerline{\includegraphics*[width=.4\textwidth,angle=0]{figs/fig_8}}
\centerline{\includegraphics*[width=\textwidth,angle=0]{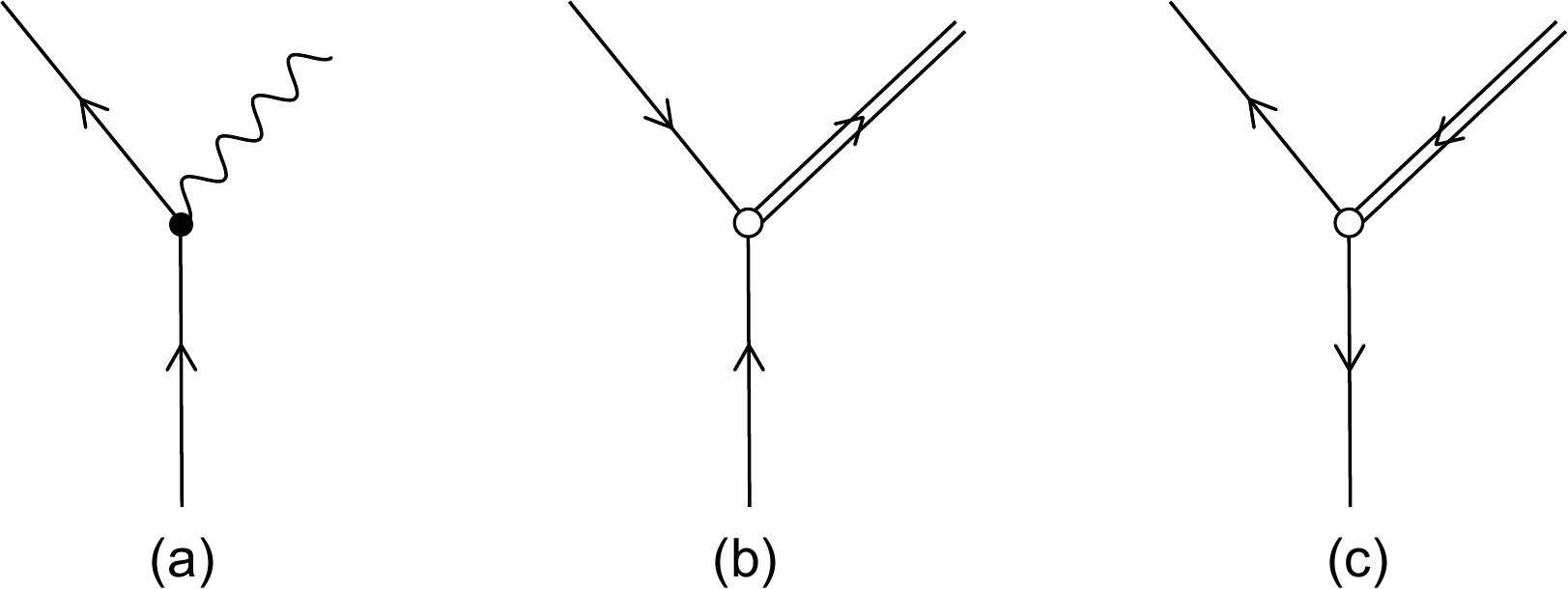}}
 \caption{Single line: particle (arrow pointing  up); hole (arrow pointing down); surface vibration (wavy  line; $\beta=0$);  
pair addition (double arrowed line, up; $beta=+2$); pair removal mode (double arrowed line, down; $\beta=-2$).}\label{fig4}
\end{figure}     
\newpage
  \begin{figure}
%\centerline{\includegraphics*[width=.4\textwidth,angle=0]{figs/fig_8}}
\centerline{\includegraphics*[width=20cm,angle=0]{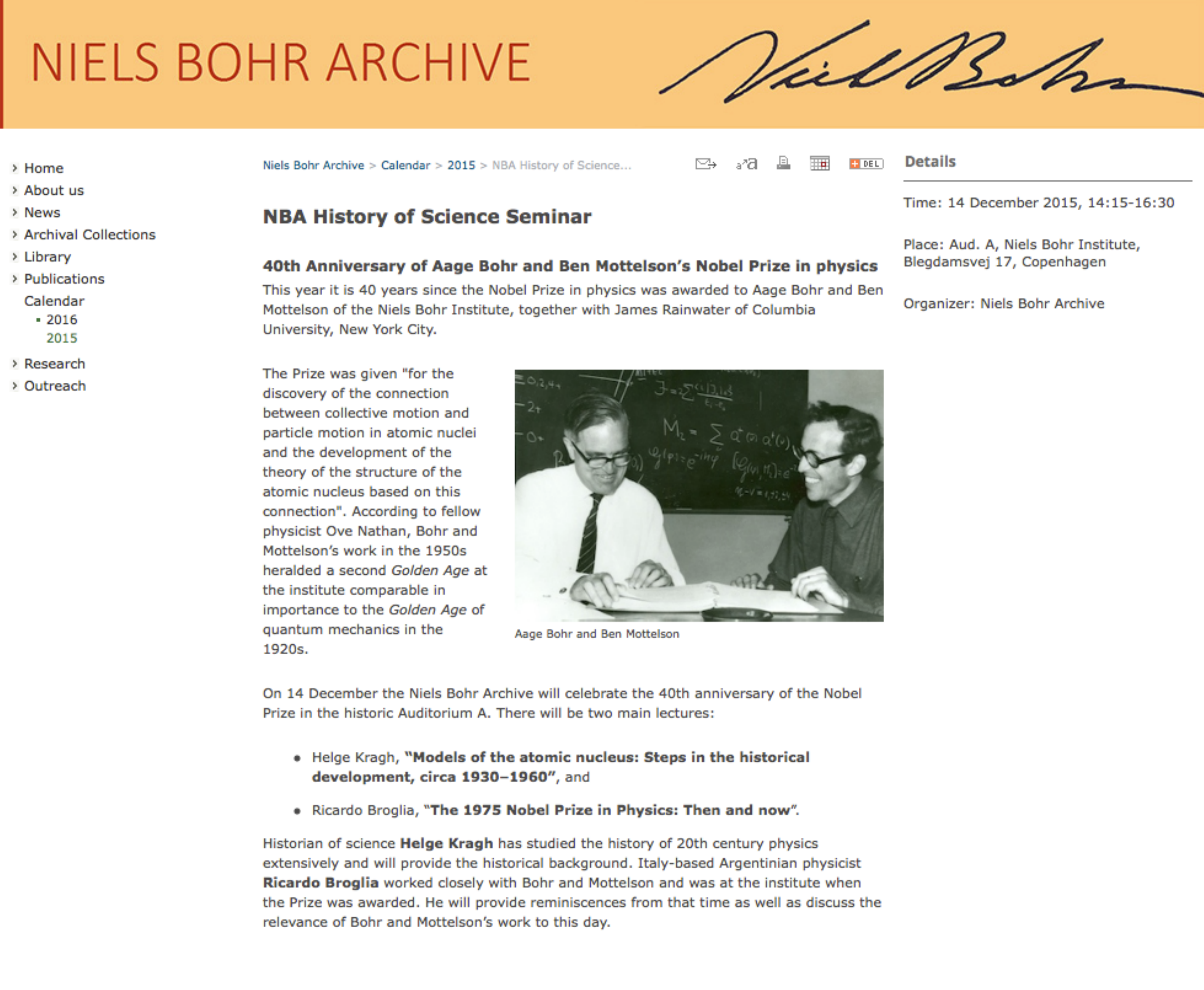}}
\end{figure} 
 \end{document}